\documentstyle[12pt]{article} 
\newcommand{\be}{\begin{equation}} 
\newcommand{\ee}{\end{equation}} 
\newcommand{\bea}{\begin{eqnarray}} 
\newcommand{\eea}{\end{eqnarray}} 
\newcommand{\nn}{\nonumber \\} 
\newcommand{\p}[1]{(\ref{#1})}

\newcommand\de{\delta}

\newcommand\ep{\epsilon}

\begin{document}
\begin{titlepage}
\begin{flushright}
hep-th/9901003 \\
January 1999
\end{flushright}

\vskip3cm
\centerline{\large\bf $N=1 \;D=4$ supermembrane in the coset approach}

\vskip2cm
\centerline{\bf E. Ivanov, S. Krivonos}

\vskip.5cm
\centerline{\it Bogoliubov Laboratory of Theoretical Physics, JINR,}
\centerline{\it 141 980, Dubna, Moscow Region, Russian Federation}

\vskip1.5cm

\begin{abstract}
\noindent $N=1\;D=4$ supermembrane (in a flat background) and 
super D2-brane dual to it are described within 
the nonlinear realizations approach as theories 
of the partial supersymmetry breaking 
$N=1\,D=4 \,\rightarrow \,N=1\, d=3$. We construct  
the relevant invariant off-shell Goldstone 
superfield actions and demonstrate them to be dual to each other. 
Their bosonic cores are, respectively, the static-gauge Nambu-Goto 
and $d=3$ Born-Infeld actions. The supermembrane 
superfield equation of motion admits  
a transparent geometric interpretation suggesting 
an extension of the standard superembedding constraints.
We briefly discuss the $1/4$ breaking of
$N=1\;D=5$ supersymmetry along similar lines.
\end{abstract}

\end{titlepage}
\noindent{\bf 1. Introduction.} The partial breaking of global 
supersymmetry (PBGS) (see, e.g. \cite{bw} - \cite{bik}) 
is a characteristic feature of supersymmetric extended objects. 
A general framework for describing PBGS is provided 
by the nonlinear realizations (coset) approach \cite{1}-\cite{3}. 
Its basic objects are Goldstone superfields 
which live on a superspace of unbroken SUSY and 
comprise a minimal worldvolume supermultiplet of given 
superbrane (a scalar one in the case of $p$-branes, a vector one in 
the case of D-branes, etc). The rest of full brane 
SUSY is realized as nonlinear transformations of Goldstone 
superfields. All the known PBGS examples correspond 
to  a ``static-gauge'' form of some BPS superbranes, 
with all their local symmetries fixed. 

One of the merits of the PBGS approach is the 
opportunity to construct manifestly 
worldvolume supersymmetric off-shell superfield actions for superbranes. 
Not too many explicit examples of such actions 
are known, and their precise relation to the Green-Schwarz-type actions 
was not fully inquired. It also remains to clear up   
how the PBGS approach is related to the superembedding one 
\cite{khark}-\cite{hs}. Note that only the $1/2$ SUSY breaking was 
treated so far. With all this in mind, it seems important 
to analyze more examples of PBGS and 
to extend its framework to the cases of $1/4, 1/8 \ldots $ 
breaking.

In this letter we describe the $N=1\;D=4$ supermembrane 
in the coset approach. It corresponds to the breaking 
$N=1\,D=4 \rightarrow N=1\, d=3$. We show that the 
basic ``geometro-dynamical'' constraint of the superembedding formalism 
\cite{{khark},{gs},{hs}} is manifested as a covariant condition on 
some Cartan 1-form, such that it expresses the Goldstone 
fermionic superfield through the basic scalar one, still leaving 
the theory off shell \cite{hs}. On the other hand, 
the equation of motion can also be given 
a nice geometric interpretation as vanishing of 
covariant spinor projection of yet another Cartan form. 
We present superfield off-shell 
PBGS actions both for the supermembrane and 
a dual super D2-brane, thus solving the long-standing problem of 
setting up such actions \cite{{town},{hrs}}. 
The D2-brane action provides a $N=2$ superextension of 
the $d=3$ Born-Infeld one. 
We also briefly discuss the 1/4 breaking of $N=1\;D=5$
SUSY. 

\vspace{0.4cm}
\noindent{\bf 2. $N=1\; D=4$ Poincar\'e superalgebra.} 
From the $d=3$ standpoint 
the $N=1\; D=4$ SUSY algebra is a central-charge 
extended $N=2$ Poincar\'e superalgebra 
\be 
N=1\;\; D=4\;\;\;\; SUSY \;\; \sim \;\;
N=2\;\; d=3\;\;\;\; SUSY
\qquad \propto \quad
\left\{ Q_a, P_{ab}, S_a, Z \right\} , \label{n1d4}
\ee
$a,b=1,2$ being the $d=3$ $SL(2,R)$ 
spinor indices \footnote{The indices are raised 
and lowered as follows:
$V^{a}=\ep^{ab}V_b,\;V_{b}=\ep_{bc}V^c,\quad
\ep_{ab}\ep^{bc}=\de_a^c\; .$}. 
The basic anticommutation relations read 
\be
\left\{ Q_{a},Q_{b}\right\}=P_{ab}\; ,\quad
\left\{ Q_{a},S_{b}\right\} = \ep_{ab}Z\; , \quad
\left\{ S_{a},S_{b}\right\} = P_{ab} \;. \label{susy}
\ee
The $d=3$ translation generator $P_{ab} = P_{ba}$ together 
with the central charge generator $Z$ form the $D=4$ translation 
generator. 

To the set \p{n1d4} one should add the generators of 
the $D=4$ Lorentz group $SO(1,3)$ consisting of the $SO(1,2)$ 
generators $M_{ab} = M_{ba}$ and the $SO(1,3)/SO(1,2)$ coset 
ones $K_{ab}=K_{ba}$
\bea
&&\left[ M_{ab},M_{cd}\right]  = \ep_{ac}M_{bd}+\ep_{ad}M_{bc}+
   \ep_{bc}M_{ad}+\ep_{bd}M_{ac} \; , \nn 
&&\left[ M_{ab},K_{cd}\right]  = \ep_{ac}K_{bd}+\ep_{ad}K_{bc}+
   \ep_{bc}K_{ad}+\ep_{bd}K_{ac} \; , \nn 
&&\left[ K_{ab},K_{cd}\right]  = -\ep_{ac}M_{bd}-\ep_{ad}M_{bc}-
   \ep_{bc}M_{ad}-\ep_{bd}M_{ac}  \; . \label{so31}
\eea
The commutation relations between the $SO(1,3)$ generators and 
those from the set \p{n1d4} read
\bea
&&\left[ M_{ab},P_{cd}\right]  = \ep_{ac}P_{bd}+\ep_{ad}P_{bc}+
   \ep_{bc}P_{ad}+\ep_{bd}P_{ac} \; , \; 
  \left[ M_{ab},Q_{c}\right]  = \ep_{ac}Q_{b}+\ep_{bc}Q_{a} \; , \nn 
&&\left[ M_{ab},S_{c}\right]  = \ep_{ac}S_{b}+\ep_{bc}S_{a} \; , \;
  \left[ P_{ab},K_{cd}\right] = 2(\ep_{ac}\ep_{bd}+\ep_{bc}\ep_{ad})Z \; , 
\nn
&&\left[ Z, K_{ab}\right] = 2P_{ab} \; , \; 
  \left[ K_{ab},Q_{c}\right]  = \ep_{ac}S_{b}+\ep_{bc}S_{a} \; , \; 
  \left[ K_{ab},S_{c}\right]  = -\ep_{ac}Q_{b}-\ep_{bc}Q_{a}  \; . 
\label{cos}
\eea 
 
\vspace{0.3cm}
\noindent{\bf 3. Coset space, transformations and Cartan forms.}
To construct a nonlinear realization of $N=1\, D=4$ SUSY (including 
the group $SO(1,3)$), such that $N=1\, d=3$ SUSY remains unbroken, 
we choose the vacuum stability subgroup to be $H  \propto 
\left\{Q_a, P_{ab}, M_{ab}\right\}$.  
We put the generators $Q_a, P_{ab}$ into the coset 
and associate with them the $N=1\; d=3$ superspace 
coordinates $\left\{ \theta^a, x^{ab} \right\}$.
The remaining coset parameters are Goldstone superfields, 
$\psi^a \equiv \psi^a(x,\theta),\;q \equiv q(x,\theta),\; \Lambda^{ab} 
\equiv \Lambda^{ab}(x,\theta)$. A coset element $g$ is defined by  
\be\label{coset}
g=e^{x^{ab}P_{ab}}e^{\theta^{a}Q_{a}}e^{qZ}
  e^{\psi^aS_a}e^{\Lambda^{ab}K_{ab}} \;.
\ee 
 
Acting on \p{coset} from the left by different elements of 
the $N=1\, D=4$ Poincar\'e supergroup, one can find 
the transformation properties of the coset coordinates.

Unbroken supersymmetry $(g_0=\mbox{exp }(a^{ab}P_{ab}+
  \eta^{a}Q_{a} ))$:
\be\label{susy1}
\de x^{ab}=a^{ab}-\frac{1}{4}\eta^a\theta^b-\frac{1}{4}\eta^b\theta^a ,
\quad  
\de \theta^{a}=\eta^a\; .
\ee

Broken supersymmetry $(g_0=\mbox{exp }(\xi^{a}S_{a}))$:
\be\label{susy2}
\de x^{ab}= -\frac{1}{4}\xi^a\psi^b-\frac{1}{4}\xi^b\psi^a,\quad
\de q=\xi^a\theta_a,\quad
\de\psi^a=\xi^a \; .
\ee

$K$ transformations $(g_0=\mbox{exp }(r^{ab}K_{ab}))$:
\bea
&&\de x^{ab}= -2q r^{ab}-\frac{1}{2}\theta_c r^{ca}\psi^b-
   \frac{1}{2}\theta_c r^{cb}\psi^a +
     \frac{1}{2}\theta^a r^{bc}\psi_c+
     \frac{1}{2}\theta^b r^{ac}\psi_c\; , \nn
&&\de \theta^{a} = -2r^{ab}\psi_b \; , 
\de q = -4r_{ab}x^{ab} \; , 
\de \psi^{ab}= 2r^{ab}\theta_b,\quad
   \de \Lambda^{ab}=r^{ab}+\ldots\quad .
   \label{ktr}
\eea

Broken $Z$-translations $(g_0 = \mbox{exp}(cZ))$: 
\be \label{Ztr}
\de q = c~.
\ee
The $d=3$ Lorentz group $SO(1,2)\sim SL(2,R)$ acts as rotations of 
the spinor indices.

As the next step of the coset formalism, one constructs the 
Cartan 1-forms 
\be
g^{-1}d g =  \Omega_Q + \Omega_P + \Omega_Z + \Omega_S + \Omega_K +
\Omega_M~, \label{cartan1} 
\ee
\bea
\Omega_Z & = & \frac{1+2\lambda^2}{1-2\lambda^2}\left[ d{\hat q}+ 
    \frac{4}{1+2\lambda^2}\lambda_{ab}d{\hat x}^{ab}\right]Z\; , \nn
\Omega_P &\equiv & \Omega_P^{ab}P_{ab} = \left[ d{\hat x}^{ab}+ 
\frac{2}{1-2\lambda^2}
\lambda^{ab}\left( 
d{\hat q} + 2 \lambda_{cd}d{\hat x}^{cd}\right) \right] P_{ab} \; ,\nn
\Omega_Q &\equiv & \Omega_Q^aQ_a = \frac{1}{\sqrt{1-2\lambda^2}}
\left[ d\theta^{a}+
     2\lambda^{ab}d\psi_{b}\right]Q_{a} \; , \;
 \Omega_S  = \frac{1}{\sqrt{1-2\lambda^2}} \left[ d\psi^{a}-
     2\lambda^{ab}d\theta_{b}\right]S_{a} \; ,\nn
 \Omega_K & = & \frac{1}{1-2\lambda^2} d\lambda^{ab} K_{ab} \; , \;
  \Omega_M  =  \frac{1}{1-2\lambda^2} \left(
   \lambda^{ac}d\lambda_c^b+\lambda^{bc}d\lambda_c^a \right) M_{ab}\; .
\label{cartan} \\
d{\hat x}^{ab}  &\equiv &  dx^{ab}+\frac{1}{4}\theta^{a}d\theta^{b}
  +\frac{1}{4}\theta^{b}d\theta^{a}+
  \frac{1}{4}\psi^a d\psi^b +\frac{1}{4}\psi^{b}d\psi^{a} \; ,\quad 
d{\hat q} \equiv  dq+\psi_{a}d\theta^{a} \; .
\eea
Here we have passed to the stereographic parametrization of 
the coset $SO(1,3)/SO(1,2)$ 
\be
\lambda^{ab}=
\frac{\mbox{th}\left(\sqrt{2\Lambda^2}\right)}{\sqrt{2\Lambda^2}}\,
\Lambda^{ab}\; ,\quad
\mbox{th}^2\left(\sqrt{2\Lambda^2}\right)\equiv 2 \lambda^2 \; ,\quad 
\Lambda^2 \equiv \Lambda_{ab}\Lambda^{ab}~, \quad 
\lambda^2 \equiv \lambda_{ab}\lambda^{ab}~.
\ee
All Cartan forms except for $\Omega_M$ are transformed 
homogeneously under all symmetries. 

In what follows, it will be convenient to deal with 
the ``semi-covariant'' derivatives 
\be
{\cal D}_{ab} =  (E^{-1})^{cd}_{ab}\,\partial_{cd} \; , \quad 
{\cal D}_a = D_a + \frac{1}{2}\psi^b D_a \psi^c \,{\cal D}_{bc} =
         D_a + \frac{1}{2}\psi^b {\cal D}_a \psi^c \,{\partial}_{bc}~,
\ee
where
\bea
&&D_a=\frac{\partial}{\partial \theta^a}+
\frac{1}{2}\theta^b\partial_{ab}\; , \quad
\left\{ D_a, D_b \right\} =\partial_{ab} \; , \label{flatd} \\
&& E_{ab}^{cd}=\frac{1}{2}(\de_a^c\de_b^d+\de_a^d\de_b^c)+
  \frac{1}{4}(\psi^c\partial_{ab}\psi^d+ \psi^d\partial_{ab}\psi^c) \;.
\eea
They obey the following algebra 
\bea
&&\left[ {\cal D}_{ab},{\cal D}_{cd} \right] =
-{\cal D}_{ab}\psi^m{\cal D}_{cd}\psi^n
           {\cal D}_{mn} \; , \nn
&&\left[ {\cal D}_{ab},{\cal D}_{c} \right] = 
{\cal D}_{ab}\psi^m{\cal D}_{c}\psi^n
           {\cal D}_{mn} \; , \nn
&&\left\{ {\cal D}_{a},{\cal D}_{b} \right\} ={\cal D}_{ab}+
        {\cal D}_{a}\psi^m{\cal D}_{b}\psi^n
           {\cal D}_{mn} \; .  \label{algebra}
\eea
These derivatives appear as the coefficients in the decompositions of 
1-superforms over the differentials $d\hat{x}^{ab}$, 
$d\theta^a$ and correspond to a truncated version of 
nonlinear realization of $N=1\;D=4$ SUSY, with only the $SO(1,2)$ subgroup 
of $SO(1,3)$ being kept (this version amounts 
to putting $\lambda^{ab} = 0$ in the above relations). The genuine 
covariant derivatives are the coefficients in the decompositions 
over the full covariant differentials $\Omega^{ab}_P$, $\Omega^a_Q$ 
defined in \p{cartan}.

\vspace{0.3cm}
\noindent{\bf 4. Kinematical and dynamical constraints.}
Not all of the above Goldstone superfields 
$\left\{ q(x,\theta),\psi^a(x,\theta), \lambda^{ab}(x,\theta)\right\}$ 
are to be treated as independent. 
Indeed, $\lambda^{ab}$ and $\psi_{a}$ appear 
inside the form $\Omega_Z$  {\it linearly} and so can be covariantly 
eliminated by the manifestly covariant constraint  
\be
\Omega_Z = 0 \label{basconstr}
\ee
(inverse Higgs effect \cite{invh}). Eq. \p{basconstr} amounts 
to the following set of equations 
\be\label{kinematik}
\mbox{(a)} \;\;{\cal D}_{ab}q+ \frac{4}{1+2\lambda^2}\lambda_{ab} = 0\; , 
\quad
\mbox{(b)}\;\;{\cal D}_a q - \psi_a = 0\; . 
\ee
Eqs. \p{kinematik} are purely algebraic nonlinear relations 
serving to express $\lambda$ and $\psi$ through $x-$ and 
$\theta$-derivatives of $q$ (the expression for $\psi$ can be obtained by 
successive iterations of eq. (\ref{kinematik}b)). 
Thus the superfield $q(x,\theta)$ is the only essential Goldstone 
superfield needed to present the partial spontaneous breaking 
$N=1\; D=4 \;\Rightarrow \; N=1\; d=3$ within the coset scheme.

The $d\theta$ part of the constraint \p{basconstr}, 
i.e. eq. (\ref{kinematik}b), 
is recognized as the ``static-gauge'' form of the 
``geometro-dynamical'' constraint \cite{{khark},{gs},{hs}} 
of the superembedding 
approach (for the flat target $N=1\;D=4$ superspace). Note 
that (\ref{kinematik}b) is covariant on its own right under 
all spontaneously broken symmetries including the coset part of 
the $D=4$ Lorentz symmetry. In agreement 
with the linearized analysis of ref. \cite{hs}, eq. \p{basconstr} 
does not imply any dynamics and leaves $q(x,\theta)$ off shell 
(as distinct, e.g., from the case treated in \cite{bik}).  

Now we want to put the additional constraint 
on the Cartan forms in order to get manifestly covariant 
dynamical equations. 
The condition we propose has no direct analogs 
in the superembedding formalism. 
We postulate the following constraint
\be\label{eom}
\Omega_S| =0 \quad \Rightarrow \quad
{\cal D}_a\psi_b+ 2\lambda_{ab} = 0 \quad \Leftrightarrow \quad 
\mbox{(a)}\;\; {\cal D}^a\psi_a = 0~, \quad \mbox{(b)}\;\;
{\cal D}_{(a}\psi_{b)}= -2\lambda_{ab}~. 
\ee
where $|$ means the ordinary $d\theta$- projection of the form $\Omega_S$.
To see that it is covariant with respect to {\it all} broken 
symmetries, it is enough to notice that, in virtue of \p{basconstr}, 
$\Omega^{ab}_P \sim A^{ab}_{cd}d\hat{x}^{cd}$ with 
$\mbox{det}A^{ab}_{cd} \neq 0$. Then the covariant 
spinor projection of $\Omega_S$ (i.e., the coefficient of $\Omega_Q^a$) 
coincides, modulo a rotation of the spinor index $a$ 
by some non-degenerate 
matrix, with the $d\theta$- projection written down in \p{eom}.  

Note that eq. (\ref{eom}a) is the consistency condition among 
eq. (\ref{eom}b) and the kinematical eqs. \p{kinematik}. 
Indeed, inserting \p{kinematik} into (\ref{eom}b) and using  
the algebra \p{algebra}, one can derive 
\be 
({\cal D}^a\psi_a)^2 = 0~. \label{consist}
\ee
It is important to realize that it the presence of the Lorentz Goldstone 
superfields $\lambda_{ab}(x,\theta)$ in the coset superspace 
which allows us to consistently write the dynamical equation 
in the geometrical form \p{eom}. Putting, e.g., $\lambda_{ab} = 0$
in \p{eom} would yield ${\cal D}_a\psi_b = 0$ that is too restrictive.

The final form of the dynamical equation is obtained by substituting 
(\ref{kinematik}b) into (\ref{eom}a)
\be  \label{fin}  
{\cal D}^a{\cal D}_a q = 0 \;.
\ee
It is recognized as a covariantization of the free equation 
of motion $D^a D_a q \equiv D^2\,q =0$. A more detailed convenient 
form of (\ref{eom}a) or \p{fin} is as follows 
\be \label{fin1}
D^a\psi_a + {1\over 2} \psi^a D^{(b}\psi^{c)}\partial_{ab}\psi_c - 
{1\over 16}\psi^2\, D^{(a}\psi^{b)} 
\partial_{ad}\psi_{f}\partial^{df}\psi_b 
= 0~.
\ee

To see, which kind of dynamics is hidden in \p{fin}, we considered it 
in the bosonic limit. We found 
that it amounts to the following equation for $q(x) \equiv 
q(x,\theta)|_{\theta = 0}$
\be  \label{NGeq}
\partial_{ab}\left( \frac{\partial^{ab}q}
       {\sqrt{1-\frac{1}{2}\partial q \cdot \partial q}}
   \right) =0~,
\ee
which corresponds to the ``static gauge'' form of the $D=4$ 
membrane Nambu-Goto action
\be  \label{NG}
S= \int d^3x \left( 1 - \sqrt{1-\frac{1}{2}\partial q \cdot \partial q} 
\right)\; .
\ee
Thus eq. (\ref{fin}) (or \p{eom}) can be naturally interpreted as a 
manifestly $N=1\; d=3$ worldvolume supersymmetric PBGS form of 
the equations of the supermembrane in $D=4$. In the next section 
we show that eq. \p{fin} can be deduced from an off-shell action.
 
\vspace{0.3cm}
\noindent{\bf 5. Superfield action of $N=1\;D=4$ supermembrane.}
The standard methods of nonlinear
realizations fail to construct the superfield action 
for the supermembrane. The full spinor and bosonic covariant
derivatives of the superfield $q(x,\theta)$, viz. $\nabla_a q$, 
$\nabla_{ab} q$, are equal to zero in virtue of 
our constraint \p{basconstr}.
Therefore, the natural candidate for the manifestly invariant action,  
$\int d^3x d^2\theta\, \mbox{sdet}E\, (\nabla^a q \nabla _{a}q)$,
identically vanishes. 

To find the action, we will follow the method of ref. \cite{bg2}.  
Let us start with a bosonic scalar superfield $\rho(x,\theta)$ 
and define the fermionic superfield
$\xi^a (x,\theta)$ 
\be \label{constr1}
\xi^a = D^a \rho \;, \qquad D^2 \xi_a = \partial_{ab} \xi^b \qquad 
(D^2 \equiv D^a D_a)~.
\ee
Let us now try to find a {\it linear} realization of an {\it extra} 
$N=1\;d=3$ SUSY on the spinor $\xi^a$ and some, arbitrary for the moment, 
scalar superfield $\Phi(x,\theta)$. Assuming the second SUSY 
to be spontaneously broken, 
the most general linear transformation law of $\xi^a$ 
can be written as
\be\label{tr1}
\delta \xi_a = \eta_a + A D^2 \Phi \eta_a + \partial_{ab} \Phi \eta^b \; ,
\ee
where $\eta_a$ is a parameter of the second SUSY and $A$ is a
constant. Requiring the standard closure of the second $N=1$ SUSY 
together with preservation of the constraint in \p{constr1}   
fixes $A=1$ and implies the following 
transformation law for the bosonic superfield $\Phi$
\be \label{tr1a}
\delta \Phi = \frac{1}{2} \eta^a \xi_a = \frac{1}{2} \eta^a D_a\rho~.
\ee

Before going further, let us make two comments.

First, the field $\Phi$ 
is a good candidate for the Lagrangian density. Indeed, the action
\be\label{action}
S=\int d^3x d^2\theta \;\Phi
\ee
is invariant as the integrand 
is shifted by a spinor derivative under the variation \p{tr1a}.

Secondly, one can extract from \p{tr1} the transformation law of the
scalar superfield $\rho$:
\be \label{tr1aa}
\delta \rho = \theta^a \eta_a - 2 D^a \Phi \eta_a \; .
\ee 
The bracket of the manifest and second $N=1$ 
SUSY's on the superfield $\rho$ yields a constant shift of $\rho$, 
in agreement with the structure relations \p{susy}. So 
in the present case we face the same 
$N=1\;\; D=4$ SUSY $\sim\;\; N=2\;\; d=3$ SUSY as before 
(the linear multiplet we have constructed is a $d=3$ 
reduction of chiral $N=1\; D=4$ multiplet).

To establish a contact with the previous consideration, 
let us first show that the superfield $\Phi $ 
can be covariantly expressed in terms of $\xi^a$. 

Close inspection
of the transformation law \p{tr1} shows that the leading part of this
superfield can be represented by $1/4 \xi^a \xi_a \equiv 1/4 \xi^2$. 
Then one finds the following recursion equation for $\Phi$:
\be\label{eqr}
\Phi = \frac{1}{4}\; \frac{ \xi^2 }{1+ D^2 \Phi} \; .
\ee
To solve it, let us note that the
nominator of \p{eqr} already contains the maximal power of the
spinors $\xi^a$. Thus the term $D^2 \Phi$ in the denominator effectively
does not contain ``free'' spinors $\xi^a$, it can contain 
only terms $D^a \xi_b$. As a result, we can write the following
equation for $(D^2 \Phi )_{eff}$:
\be\label{phieq}
(D^2 \Phi )_{eff} =\frac{1}{4}\; 
\frac{ D^2 \xi^2 }{1+ (D^2 \Phi)_{eff}} \; .
\ee
Solving eq. \p{phieq} and specializing to the solution which goes to
zero in the limit $\xi \rightarrow 0$, we find 
\be\label{sol1}
(D^2 \Phi)_{eff} = -\frac{1}{2}\left(1- \sqrt{1+D^2 \xi^2}\right) 
\quad \Rightarrow \quad 
\Phi =  \frac{1}{2}\; \frac{ \xi^2}{1+\sqrt{1+D^2 \xi^2}} \; .
\ee
Finally, the action \p{action} takes  the
following form
\be\label{action1}
S= \int d^3 x d^2 \theta\, \Phi \equiv \frac{1}{2}  
\int d^3 x d^2 \theta
\, \frac{ \xi^2}{1+\sqrt{1+D^2\xi^2}} \;, \quad 
\xi^a = D^a\,\rho~.
\ee
For the physical bosonic component $\rho|_{\theta=0}$ 
one obtains just the Nambu-Goto action \p{NG}.

Now we are ready to reveal the relation with 
the previously used superfields $\psi^a$ and $q$. 
One can check that under the transformations \p{tr1}, 
\p{tr1a} the object 
\be \label{equiv}
\psi^a = {\xi^a\over 1 + D^2\Phi}
\ee
transforms according to the law 
\be
\delta \psi^a = \eta^a+ \frac{1}{2} \eta^b\psi^c\partial_{bc}\psi^a~. 
\ee
This is just the ``active'' form of the transformations \p{susy2}, 
so $\psi^a$ in \p{equiv} can be identified with 
the nonlinear realizations 
Goldstone fermionic superfield. After a straightforward, 
though cumbersome computation with making use of the equivalence 
relation \p{equiv}, the equation of motion 
corresponding to \p{action1} can be written 
in terms of $\psi^a$ as  
\be  \label{eqm2}
D^a\left[
{\psi_a \over 1 - {1\over 2}(D\psi)^2} 
+{1\over 4}{\psi^2 \over [1 - {1\over 2}(D\psi)^2]^2} \left( 
\partial_{an}\psi^n +2D_l\psi_a D_m\psi_c \partial^{lm}\psi^c 
\right)\right] = 0~.
\ee
Although, at first sight, \p{eqm2} radically differs from the 
previously conjectured eq. \p{fin1}, we have checked that 
the former implies the latter and vise versa! The proof 
\cite{dik} is based, first, on the fact that 
eq. \p{eqm2} encodes eq. \p{fin1} together with some corollaries thereof 
and, second, that $\psi^a$ satisfies 
a nonlinear version of the kinematical irreducibility constraint 
\p{constr1} (it can be deduced using either eq.\p{equiv} or 
(\ref{kinematik}b)). Thus, eq.\p{fin1} conjectured on 
the purely geometric grounds amounts to eq. \p{eqm2} 
obtained from the off-shell action \p{action1}. This one-to-one 
correspondence confirms the self-consistency of our approach and 
the relevancy of the extension \p{eom} of the standard 
superembedding postulate. It also implies that \p{action1} 
possesses all the symmetries of eq. \p{fin1}, including 
$SO(1,3)$ symmetry. Note that one can find the 
equivalence relation directly between the unconstrained scalar 
superfields $\rho(x,\theta)$ and $q(x,\theta)$ and rewrite 
\p{action1} via $q(x,\theta)$. When all fermions are discarded, 
the relation $\partial_{ab}q = \partial_{ab}\rho $ follows. 
It means that the physical bosonic fields 
in eqs. \p{NG}, \p{NGeq} and in the action \p{action1} 
coincide up to nilpotent additions.

\vspace{0.3cm}
\noindent{\bf 6. D2-brane.} 
Like in other PBGS theories, in our case 
the Goldstone fermion can be placed into different multiplets 
of unbroken $N=1\; d=3$ SUSY. Besides a scalar multiplet,  
we can choose a vector multiplet as the basic Goldstone one. 
In a field-strength formulation it is represented by $N=1$ 
spinor superfield $\mu_a$ subjected to the constraint \cite{bibl}:
\be\label{cc1}
D^a\mu_a=0 \quad \Rightarrow \quad 
\left\{
 \begin{array}{l}
   D^2 \mu_a=-\partial_{ab}\mu^b  \nn
   \partial_{ab}D^a\mu^b = 0~. \nonumber
  \end{array} \right. 
\ee
It leaves in $\mu_a$ the first fermionic 
(Goldstone) component together with the 
divergenceless vector $F_{ab}\equiv D_a\mu_b|_{\theta=0}$ 
(just the gauge field strength). Due to the vector-scalar $d=3$ duality, 
the superfield $\mu_a$ is expected to describe 
a D2-brane which is dual to the supermembrane.

In constructing the relevant Goldstone superfield action we 
again follow ref. \cite{bg2}. We start by defining 
a {\it linear} realization of the second
(broken) SUSY on the superfield $\mu^a$ and some 
scalar superfield $\phi$. The unique possibility consistent 
with the constraint \p{cc1} is \footnote{This is 
a $d=3$ reduction of $N=1\; D=4$ linear (or tensor) 
multiplet \cite{bibl}.}.
\be \label{tr2}
\delta \mu_a = \eta_a - D^2 \phi \eta_a + \partial_{ab} \phi \eta^b \; ,
\qquad
\delta \phi = \frac{1}{2} \eta^a \mu_a \;.
\ee
Like in the supermembrane case, one can write the recursion equation
\be
\phi= \frac{1}{4}\; \frac{\mu^2 }{1-D^2\phi}
\ee
and solve it to get
\be
\phi= \frac{1}{2}\; \frac{\mu^2 }{1+\sqrt{1-D^2\mu^2}} \; .
\ee
Due to the transformation law of $\phi$ \p{tr2} and the basic constraint
\p{cc1}, the action
\be\label{actionm}
S=-\int d^3x d^2\theta\, \phi = - \frac{1}{2}\,\int d^3x d^2\theta
\,\frac{\mu^2 }{1+\sqrt{1-D^2\mu^2}} 
\ee
is invariant under the second SUSY \p{tr2}. Its bosonic core is the 
$d=3$ Born-Infeld action
\be\label{bm}
S= \int d^3x \left( \sqrt{1+2F^2}-1 \right) \;,
\ee
where 
\be
\partial^{ab}F_{ab}=0 \quad \rightarrow \quad
     F_{ab}=\partial_{ac}G^c_b+\partial_{bc}G^c_a \; .
\ee

To find how the action \p{actionm} is related to the 
supermembrane one \p{action1} we make  
the duality transformation. We add the constraint \p{cc1} 
to \p{actionm} with a superfield Lagrange multiplier:
\be\label{a2}
S\rightarrow   -{1\over 2} \int d^3x d^2\theta
\left( \frac{\mu^2 }{1+\sqrt{1-D^2\mu^2}}+ D_a\rho \mu^a \right) \;.
\ee 
Varying \p{a2} with respect to the unconstrained superfield $\mu_a$, 
we find
\be\label{xi2}
\xi_a\equiv D_a\rho= \frac{2\,\mu_a}{1+\sqrt{1-D^2\mu^2}}+
 \mu_a
  D^2\left[ \frac{\mu^2}{\left( 1+\sqrt{1-D^2\mu^2}\right)^2 
\sqrt{1-D^2\mu^2}}
  \right] \;.
\ee
We then substitute \p{xi2} back into \p{a2}
\be\label{a3}
S=\frac{1}{2}\int d^3x d^2\theta
\,\frac{\mu^2}{\left( 1+\sqrt{1-D^2\mu^2}\right) \sqrt{1-D^2\mu^2}}=
\frac{1}{8}\,\int d^3x d^2\theta\,\frac{\xi^2}{X} \; ,
\ee
where
\be
X= \frac{1}{1+\sqrt{1-D^2\mu^2}}+\frac{1}{4}\,
\frac{\left(D^2\mu^2\right)^2}
   {\left( 1+\sqrt{1-D^2\mu^2}\right)^3\sqrt{1-D^2\mu^2}} \;.
\ee
The last step is to express $X$ in terms of $\xi^a$. We
need to know only the ``effective'' form of $X$ because 
the nominator \p{a3}
already contains the maximal power of $\xi$. So, from
\p{xi2} one gets
\be\label{xi3}
\frac{1}{4}D^2 \xi^2 \propto 
  \frac{D^2\mu^2}{\left( 1+\sqrt{1-D^2\mu^2})\right)
  \sqrt{1-D^2\mu^2}}X \;,
\ee
where ``$\propto$'' means that the equation is valid up to 
spinors with no derivatives. It is easy to find 
\be
\frac{D^2\mu^2}{\left( 1+\sqrt{1-D^2\mu^2})\right)
  \sqrt{1-D^2\mu^2}}=4X-2
\ee
and hence
\be \label{quadr}
\frac{1}{4}D^2 \xi^2 \propto (4X-2)X \;.
\ee
Solving eq. \p{quadr} for $X$, we finally find 
\p{a3} to be just the supermembrane action \p{action1}.

Thus we have demonstrated that the action \p{actionm} is dual to the
supermembrane action, possesses partially broken $N=2$ SUSY and
is reduced to the Born-Infeld action in the bosonic limit. 
So, we may conclude that \p{actionm} is a gauge-fixed 
super D2-brane action or, equivalently, $N=1$ superextension of the 
$d=3$ Born-Infeld action with a hidden nonlinearly realized 
second SUSY. We did not study how to reproduce this 
$N=2 \;\Rightarrow \;N=1$ PBGS pattern directly 
from the nonlinear realizations. 
The duality of the action \p{actionm} to the supermembrane one is an 
indication that the former possesses all symmetries 
of the latter. This suggests that in the present case one should start 
from a nonlinear realization of the same $N=1\; D=4$ supergroup, 
but place the generator $Z$ into the vacuum stability subgroup 
in order to avoid the presence of scalar Goldstone superfield 
in the coset. 

\vspace{0.3cm}
\noindent{\bf 7. $N=4\rightarrow N=1$ breaking.}
Finally, we briefly discuss the $1/4$ partial breaking 
$N=1\;D=5\;\;(N=4\;d=3) \;\rightarrow \; N=1\; d=3$ along 
the lines of previous sections. 

From the $d=3$ point of view, 
extending $N=1\;D=4$ SUSY to $N=1\; D=5$ amounts to adding one more 
translation generator $P_{5}$, two real supertranslation 
generators  ${\bar Q}_a, {\bar S}_a$ and two extra Lorentz generators 
$L_{ab}$ and $U$ which complement $SO(1,3)$ to $SO(1,4)$. We 
as before wish  
$N=1\; d=3$ SUSY to be unbroken, so we are led to add new Goldstone 
superfields
\be
P_{5}\; \Rightarrow \;{\bar q}(x,\theta)~, \;\; {\bar Q}_a \;\Rightarrow 
\; \xi^a(x,\theta)~, \;\; {\bar S}_a \;\Rightarrow\; 
{\bar\psi}^a(x,\theta)~, \;\;
L_{ab} \;\Rightarrow \; u^{ab}(x,\theta)~,\;\; 
U \;\Rightarrow \;u(x,\theta)~. \label{newG}
\ee
The linearized analysis shows that 
${\bar\psi}^a(x,\theta),u^{ab}(x,\theta)$ and
$u(x,\theta)$ can be covariantly expressed in terms of 
$ {\bar q}(x,\theta)$ and
$\xi^a(x,\theta)$. Thus, the set of unremovable Goldstone superfields 
enlarges to $\{ q, {\bar q}, \xi^a \}$. The superfield $\xi^a$ 
is reducible, and we impose a proper constraint on it (cf. eq. \p{cc1}):
\be\label{nc1}
D^a\xi_a =0~. 
\ee
$N=4\;d=3$ SUSY should be {\it nonlinearly} realized on 
this minimal set of Goldstone superfields.

By analogy with the supermembrane and D2-brane, we can try to extend
this set by an additional scalar bosonic superfield $\Phi$ to
achieve a {\it linear } realization of $N=4$ SUSY on 
the enlarged set. Curiously, such an extension exists and 
the resulting transformations mimic those for 
the membrane \p{tr1}-\p{tr1aa} and D2-brane \p{tr2}.

\vspace{0.1cm}
\noindent{Second supersymmetry:}
\bea\label{n41}
&& \delta\xi_a=\nu_a - D^2\Phi\nu_a+\partial_{ab}\Phi\nu^b \; , \;
   \delta \Phi = \frac{1}{2} \nu^a\xi_a \; , \nn
&& \delta q = \nu^a D_a {\bar q} \; , \;
   \delta{\bar q} = -\nu^a D_a q \:.
\eea
\noindent{Third supersymmetry:}
\bea\label{n42}
&& \delta\xi_a = -D^2 q\epsilon_a+\partial_{ab}q \epsilon^b \; , \;
    \delta q = -\epsilon^a\xi_a \; , \nn
&& \delta {\bar q} = \theta^a\epsilon_a -2 D^a\Phi\epsilon_a \; , \;
   \delta\Phi =-\epsilon^a D_a{\bar q} \; .
\eea 
\noindent{Fourth supersymmetry:}
\bea\label{n43}
&& \delta\xi_a = D^2 {\bar q}{\bar\epsilon}_a-\partial_{ab}{\bar q} 
{\bar\epsilon}{}^b\; , \;
\delta {\bar q} = {\bar\epsilon}{}^a\xi_a \; , \nn
&& \delta q = \theta^a{\bar\epsilon}_a - 
2 D^a\Phi{\bar\epsilon}_a \; , \;
   \delta\Phi =-{\bar\epsilon}{}^a D_a q \; .
\eea 
Here, $\nu_a, \epsilon_a$ and ${\bar\epsilon}_a$ are the 
corresponding parameters. This representation is seemingly a $d=3$ form 
of the $D=5$ version of $N=2\; D=4$ linear (tensor) multiplet \cite{tens}.

As in the previously considered cases, the simplest action
\be\label{alast}
S=\int d^3x d^2\theta\; \Phi
\ee
is invariant under $N=4\; d=3$ SUSY, 
and our task is to express the superfield $\Phi$ 
in terms of $q,\bar q,\xi_a$. The simplest way to do this is to start
from the ansatz
$$
\Phi=\frac{1}{4}\xi^a\xi_a-\frac{1}{2}D^aqD_aq-
\frac{1}{2}D^a{\bar q}D_a{\bar q}+
    \mbox{ higher order terms}~,
$$
and to find higher-order terms by requiring $\delta\Phi$ 
to be as in eqs. \p{n41}-\p{n43}. The resulting action is uniquely 
determined and, up to the third order in fields, reads
\bea
S&=&\int d^3x d^2\theta\left[ \frac{1}{4}\xi^a\xi_a-\frac{1}{2}D^aqD_aq-
  \frac{1}{2}D^a{\bar q}D_a{\bar q}+\xi^b\left(D^a qD_aD_b{\bar q}-
    D^a{\bar q}D_aD_b q\right) \right. \nn
  & & \left. -D^a{\bar q}D^bqD_a\xi_b +\ldots \right]~.
\eea

\vspace{0.3cm}
\noindent{\bf 8. Concluding remarks.}
In this paper we have constructed, for the first time, 
off-shell manifestly $d=3$ supervolume supersymmetric PBGS 
actions for $N=1\;D=4$ supermembrane and for its dual 
super D2-brane (in a flat background). The former action is 
expected to provide an off-shell 
superfield form of the component on-shell action found in \cite{town} by 
fixing the static gauge in the $D=4$ supermembrane Green-Schwarz action. 
The D2-brane action is an $N=2$ superextension 
of the $d=3$ Born-Infeld action with a nonlinearly realized second SUSY.
The supermembrane superfield equation of motion 
was shown to have a geometric interpretation of vanishing of 
the odd supervolume projection of the covariant differential of 
Goldstone fermionic superfield. This 
implies an extension of the standard ``geometro-dynamical'' 
postulate of the superembedding 
approach. Actually, it matches with the ``double-analyticity'' 
principle put forward in early works on this approach 
\cite{iak1}, being its ``extreme'' dynamical form. It would 
be of interest to find further examples to which such an 
extension is relevant. Also, it would be tempting to 
establish a relation with the $D=4$ supermembrane 
worldvolume action proposed in \cite{hrs} within the superembedding 
formalism. One more 
(perhaps, most intriguing) problem for the future study  is to 
further elaborate on the description of the $1/4, 1/8, ... $ PBGS  
in the framework of nonlinear realizations, 
with the toy model of sect.7 as a starting point. 
 
\vspace{0.4cm}
\noindent{\bf Acknowledgments.} We thank I. Bandos, L. Bonora, P. Pasti, 
D. Sorokin, M. Tonin and, especially, F. Delduc for their interest 
in the work and useful comments. 
E.I. and S.K. acknowledge the partial 
support from the grants RFBR 96-02-17634, RFBR-DFG 98-02-00180, 
RFBR-CNRS 98-02-22034, INTAS-93-127ext, INTAS-96-0538 and INTAS-96-0308. 
S.K. thanks ENS (Lyon) for the hospitality under the PICS Project N 593.

\end{document}